\documentclass[10pt,conference]{IEEEtran}
\usepackage{amssymb}
\usepackage{amsxtra}
\usepackage{amsmath}
\usepackage{latexsym}
\usepackage{graphics}
\usepackage{verbatim}
\usepackage{epsfig}
\usepackage{setspace}

\newtheorem{theorem}{Theorem}[section]
\newtheorem{lemma}[theorem]{Lemma}

\newtheorem{definition}{Definition}[section]
\newtheorem{example}[theorem]{Example}

\def\bA{\textbf{A}}
\def\bC{\textbf{C}}
\def\bG{\textbf{G}}
\def\bT{\textbf{T}}
\def\bU{\textbf{U}}
\def\bq{\textbf{q}}
\def\bp{\textbf{p}}

\def\p{{\mathbf p}}
\def\q{{\mathbf q}}
\def\r{{\mathbf r}}
\def\x{{\mathbf x}}
\def\y{{\mathbf y}}

\def\bar{\overline}

\def\cC{{\cal C}}

\begin{document}

\title{DNA Codes that Avoid Secondary Structures}

\author{\authorblockN{Olgica Milenkovic}
\authorblockA{Dept.\ Electrical and Computer Eng.\\
University of Colorado \\
Boulder, CO, 80309, USA. \\
Email: \texttt{olgica.milenkovic@colorado.edu}}
\and
\authorblockN{Navin Kashyap}
\authorblockA{Dept.\ Mathematics and Statistics \\
Queen's University \\
Kingston, ON K7L3N6, Canada. \\
Email: \texttt{nkashyap@mast.queensu.ca}}
}
%\author{Olgica Milenkovic \\
%{\small University of Colorado} \\
%{\small Boulder, CO 80309, USA.} \\
%{\small \texttt{olgica.milenkovic@colorado.edu}} \and
%Navin Kashyap \\
%{\small Queen's University} \\
%{\small Kingston, ON, K7L 3N6, Canada.} \\
%{\small\texttt{nkashyap@mast.queensu.ca}}}
\date{}
\maketitle

\footnotetext{This work was supported in part by a research grant 
from the Natural Sciences and Engineering Research Council (NSERC) of Canada.}

\renewcommand{\thefootnote}{\arabic{footnote}}
\setcounter{footnote}{0}

\begin{abstract}
In this paper, we consider the problem of
designing codewords for DNA storage systems and DNA computers
that are unlikely to fold back onto themselves to form undesirable
secondary structures. Secondary structure formation causes a DNA
codeword to become less active chemically, thus rendering it useless
for the purpose of DNA computing. It also defeats the read-back
mechanism in a DNA storage system, so that information stored
in such a folded DNA codeword cannot be retrieved.
Based on some simple properties of a dynamic-programming algorithm,
known as Nussinov's method, which is an effective predictor of
secondary structure given the sequence of bases in a DNA codeword,
we identify some design criteria that reduce the possibility of
secondary structure formation in a codeword. These design criteria
can be formulated in terms of the requirement that the
Watson-Crick distance between a DNA codeword and a number of its
shifts be larger than a given threshold.
%, or that the DNA codewords
%do not contain long subsequences and their reverse complements.
This paper addresses both the issue of enumerating DNA sequences
with such properties and the problem of practical DNA code construction.
\end{abstract}

%%%%%%%%%%%%%%%%%%%%%%%%%%%%%%%%%%%%%%%%%%%%%%%%%%%%%%%%%%%%%%%%%%%%%%%%%%%%%%

\section{Introduction}

The last century was marked by the birth of two major
scientific and engineering disciplines: silicon-based computing
and the theory and technology of genetic data analysis. The
research field very likely to dominate the area of scientific
computing in the foreseeable future is the merger of these two
disciplines, leading to unprecedented possibilities for
applications in varied areas of engineering and science. The first
steps toward this goal were made in 1994, when Leonard Adleman
\cite{Adleman94} solved a quite unremarkable computational problem,
an instance of the directed travelling
salesmen problem on a graph with seven nodes,
with an exceptional method. The technique
used for solving the problem was a new technological paradigm,
termed DNA computing. DNA computing introduced the possibility of
using genetic data to tackle computationally hard classes of
problems that are otherwise impossible to solve using traditional
computing methods. The way in which DNA computers make it possible
to achieve this goal is through massive parallelism of operation on
nano-scale, low-power, molecular hardware and software systems.

One of the major obstacles to efficient DNA computing, and more
generally DNA storage \cite{masud} and signal processing
\cite{sotirios}, is the very low reliability of single-stranded
DNA sequence operations. DNA computing experiments require the
creation of a controlled environment that allows for a set of
single-stranded DNA codewords to bind (hybridize) with their
complements in an appropriate fashion. If the codewords are not
carefully chosen, unwanted, or non-selective, hybridization may
occur. Even more detrimental is the fact that a single-stranded
DNA sequence may fold back onto itself, forming a secondary
structure which completely inhibits the sequence from
participating in the computational process. Secondary structure
formation is also a major bottleneck in DNA storage systems. For
example, it was reported in \cite{masud} that $30\%$ of read-out
attempts in a DNA storage system failed due to the formation of
special secondary structures called \emph{hairpins} in the
single-stranded DNA molecules used to store information. 

So far, the focus of coding for DNA computing was exclusively directed
towards constructing large sets of DNA codewords with fixed
base frequencies (constant GC-content) 
and prescribed Hamming/reverse-complement Hamming
distance. Such sets of codewords are expected to result in very
rare hybridization error events. As an example, it was shown in
\cite{gaborit} that there exist $94595072$ codewords of length
$20$ with minimum Hamming distance $d=5$ and with exactly $10$
\bG/\bC$\,$ bases. At the same time, the Wisconsin DNA
Group, led by D.\ Shoemaker, reported that through extensive
computer search and experimental testing, only $9105$
sequences of length $20$ at Hamming distance at least $5$ were
found to be free of secondary structure at temperatures of $61 \pm
5^{o}C$. Since at lower ambient temperatures the probability of
secondary structure formation is even higher, it is clear that
the effective number of codewords useful for DNA computing is
extremely small.

%In a companion paper \cite{milkas}, we proposed several simple
%code construction techniques that allow for efficient testing for
%possible secondary structure formations. 
In this paper, we investigate properties of DNA sequences that
may lead to undesirable folding. Our approach is based on
analysis of a well-known algorithm for approximately determining
DNA secondary structure, called Nussinov's method. This analysis
allows us to extract some design
criteria that yield DNA sequences that are unlikely to fold
undesirably. These criteria reduce to the requirement that the
first few shifts of a DNA codeword have the property that they do
not contain Watson-Crick complementary matchings with the original
sequence.
%and, that the DNA sequences do not contain long subsequences and
%their reverse complements at distances within some given range
%$[D_1,D_2]$.
We consider the enumeration of sequences having the shift property
and provide some simple construction strategies which
meet the required restrictions.
%For the second constraint, we
%outline a set of construction methods that can be used in this
%setting.
To the best of our knowledge, this is the first attempt
in the literature aimed at providing a rigorous setting that links
DNA folding properties to constraints on the primary structure of the 
sequences.

\section{DNA Secondary Structure: Properties and Code Design Issues}

DNA of higher species consists of two complementary chains twisted
around each other to form a double helix. Each chain is a linear
sequence of nucleotides, or \emph{bases}--- two \emph{purines},
adenine (\bA) and guanine (\bG), and two \emph{pyrimidines},
thymine (\bT) and cytosine (\bC). The purine bases and pyrimindine
bases are \emph{Watson-Crick (WC) complements} of each other, in the
sense that
\begin{equation}
\bar{\bA} = \bT, \ \ \ \bar{\bG} = \bC, \ \ \ \bar{\bC} = \bG, \
\ \ \bar{\bT} = \bA.
\label{bar_def}
\end{equation}
More specifically, in a double helix, the
base \bA\ pairs with \bT\ by means of two hydrogen bonds, while
\bC\ pairs with \bG\ by means of three hydrogen bonds (i.e. the
strength of the former bond is weaker than the strength of the
latter). For DNA computing purposes, one is only concerned with
single-stranded (henceforth, \emph{oligonucleotide}) DNA sequences.

Oligonucleotide DNA sequences are formed by heating DNA double helices to
denaturation temperatures, at which they break down into single
strands. If the temperature is subsequently reduced,
oligonucleotide strands with large regions of sequence
complementarity can bind back together in a process
called \emph{hybridization}.
Hybridization is assumed to occur only between complementary base
pairs, and lies at the core of DNA computing.

As a first approximation, oligonucleotide DNA sequences can be
simply viewed as words over a four-letter alphabet $Q =
\{\bA,\bC,\bG,\bT\}$, with a prescribed set of complex properties.
The generic notation for such sequences will be $\mathbf{q} =q_1
q_2 \ldots q_n$, with $n$ indicating the length of the sequences.
The WC complement $\bar{\bq}$ of a DNA sequence is
defined as $\bar{q_1} \, \bar{q_2} \ldots \bar{q_n}$, $\bar{q_i}$ being
the WC complement of $\bar{q_i}$ as given by (\ref{bar_def}).

\begin{figure}
\begin{center}
\epsfig{file=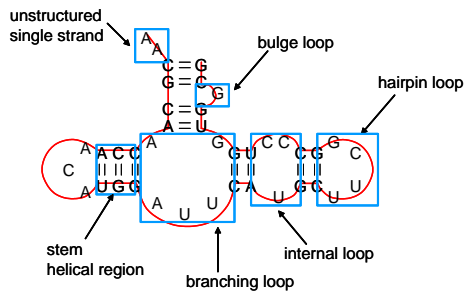, bb=0 220 612 572, clip=, width=8cm}
\caption{\label{sec-str} DNA/RNA secondary structure model
(reprinted from~\cite{mneimneh}).}
\end{center}
\end{figure}

The \emph{secondary structure} of a DNA codeword $q_1q_2 \ldots
q_n$ is a set, $S$, of disjoint pairings between complementary
bases $(q_i,q_j)$ with $i < j$. A secondary structure is formed by
a chemically active oligonucleotide sequence folding back onto
itself due to \emph{self-hybridization}, \emph{i.e.},
hybridization between complementary base pairs belonging to the
same sequence. As a consequence of the bending, elaborate spatial
structures are formed, the most important components of which are
loops (including branching, internal, hairpin and bulge loops),
stem helical regions, as well as unstructured single strands.
Figure~\ref{sec-str} illustrates these concepts for an RNA
strand\footnote{Oligonucleotide DNA sequences are structurally
very similar to RNA sequences, which are by their very nature
single-stranded, and consist of the same bases as DNA strands,
except for thymine being replaced by uracil (\bU).}. It was shown
experimentally that the most important factors influencing the
secondary structure of a DNA sequence are the number of base pairs
in stem regions, the number of base pairs in a hairpin loop region
as well as the number of unpaired bases.
%for a DNA sequence of the form
%$$
%AACGCAACCAACATGGATTCATGCTTCGGGCCCTGGTGGCG.$$

Determining the exact pairings in a secondary structure of a DNA
sequence is a complicated task, as we shall try to explain briefly.
For a system of interacting entities, one measure commonly used
for assessing the system's property is the free energy.
The stability and form of a secondary configuration is usually
governed by this energy, the general rule-of-thumb being that a secondary
structure minimizes the free energy associated with a DNA sequence.
The free energy of a secondary structure is determined by the
energy of its constituent pairings.
Now, the energy of a pairing depends on the bases
involved in the pairing as well as all bases adjacent to it.
Adding complication is the fact that in the presence of
other neighboring pairings,
these energies change according to some nontrivial rules.

Nevertheless, some simple dynamic programming techniques can be
used to \emph{approximately} determine the secondary structure of
a DNA sequence. Such approximations usually have the correct form
in $70\%$ of the cases considered. Among these techniques,
\emph{Nussinov's folding algorithm} is the most widely used scheme
\cite{nussinov}. Nussinov's algorithm is based on the assumption
that in a DNA sequence $c_1 c_2 \ldots c_n$, the energy between a
pair of bases, $\alpha(c_i,c_j)$, is independent of all other
pairs. For simplicity of exposition, we shall assume that
$\alpha(c_i,c_j)=-1$ if $\{{c_i,c_j\}}=\{{\bA,\bT\}}$,
$\alpha(c_i,c_j)=-2$ if $\{{c_i,c_j\}}=\{{\bG,\bC\}}$, and
$\alpha(c_i,c_j)=0$ otherwise\footnote{Experimentally obtained
interaction energies, depending on the choice of the base pair,
can be easily incorporated into the model instead of the constants
$-1$ and $-2$.}. Let $E_{i,j}$ denote the minimum free energy of the
subsequence $c_i,\ldots,c_j$. The independence assumption allows
us to compute the minimum free energy of the sequence
$c_1,c_2,\ldots,c_n$ through the recursion
\begin{equation} \label{free-energy}
E_{i,j}=\min \left\{\begin{array}{cc}
  E_{i+1,j-1}+\alpha(c_i,c_j), &  \\
  E_{i,k-1}+E_{k,j}, & i<k\leq j,  \\
\end{array} \right.
\end{equation}
where $E_{i,i}=0$ for $i=1,2,...,n$ and $E_{i,i-1}=0$ for
$i=2,...,n$. The value of $E_{1,n}$ is the minimum free energy of
a secondary structure of $c_1,c_2,...,c_n$.
Note that $E_{1,n} \leq 0$.
A very large negative value for the free energy
$E_{1,n}$ of a sequence is a good indicator of the
presence of stacked base pairs and loops, \emph{i.e.},
a secondary structure, in the physical
DNA sequence.

%The complexity of Nussinov's folding algorithm is of the order of
%$O(n^3)$, while the complexity of the backtracking algorithm is
%$O(n)$. The algorithm is part of many commercial secondary
%structure prediction programs, such as the \emph{Vienna secondary
%structure package} \cite{vienna}. The Vienna package uses more
%accurate values of the parameters $\alpha(c_i,c_j)$, and it has an
%additional option for predicting the base pairing probabilities.

Nussinov's algorithm can be described in terms of free-energy
tables, two of which are shown below. We first describe how such
a table is filled out, after which we will point out some
important properties of such tables.
\begin{table}[hbt]
\centering
\begin{tabular}{cccccccccc}
& \textbf{G} & \textbf{C} & \textbf{G} & \textbf{C} & \textbf{C} &
     \textbf{C} & \textbf{C} & \textbf{G} & \textbf{C} \\
\textbf{G} & \underline{0} & -2 & -2 & -4 & -4 & -4 & -4 & -6 & -6 \\
\textbf{C} & \underline{0} & \underline{0} & -2 & -2 & -2 & -2 & -2 & -4 & -4 \\
\textbf{G} & * & \underline{0} & \underline{0} & -2 & -2 & -2 & -2 & -4 & -4 \\
\textbf{C} & * & * & \underline{0} & \underline{0} & 0 & 0 & 0 & -2 & -2 \\
\textbf{C} & * & * & * & \underline{0} & \underline{0} & 0 & 0 & -2 & -2 \\
\textbf{C} & * & * & * & * & \underline{0} & \underline{0} & 0 & -2 & -2 \\
\textbf{C} & * & * & * & * & * & \underline{0} & \underline{0} & -2 & -2 \\
\textbf{G} & * & * & * & * & * & * & \underline{0} & \underline{0} & -2 \\
\textbf{C} & * & * & * & * & * & * &  * & \underline{0} & \underline{0} \\
\end{tabular}
\caption{Free-energy table for the sequence \bG\bC\bG\bC\bC\bC\bC\bG\bC}
\label{table1}
\end{table}
\begin{table}[tbh]
\centering
\begin{tabular}{cccccccccc}
& \textbf{G} & \textbf{A} & \textbf{G} & \textbf{G} &
      \textbf{G} & \textbf{T} & \textbf{T} & \textbf{T} & \textbf{T} \\
\textbf{G} & \underline{0} & 0 & 0 & 0 & 0 & -1 & -1 & -1 & -1 \\
\textbf{A} & \underline{0} & \underline{0} & 0 & 0 & 0 & -1 & -1 & -1 & -1 \\
\textbf{G} & * & \underline{0} & \underline{0} & 0 & 0 & 0 & 0 & 0 & 0 \\
\textbf{G} & * & * & \underline{0} & \underline{0} & 0 & 0 & 0 & 0 & 0 \\
\textbf{G} & * & * & * & \underline{0} & \underline{0} & 0 & 0 & 0 & 0 \\
\textbf{T} & * & * & * & * & \underline{0} & \underline{0} & 0 & 0 & 0 \\
\textbf{T} & * & * & * & * & * & \underline{0} & \underline{0} & 0 & 0 \\
\textbf{T} & * & * & * & * & * & * & \underline{0} & \underline{0} & 0 \\
\textbf{T} & * & * & * & * & * & * & * & \underline{0} & \underline{0} \\
\end{tabular}
\caption{Free-energy table for the sequence \bG\bA\bG\bG\bG\bT\bT\bT\bT}
\label{table2}
\end{table}
In a free-energy table, the entry at position $(i,j)$ (the top left position
being (1,1)), contains the value of $E_{i,j}$. The table is
filled out by initializing the entries on the main diagonal and on
the first lower sub-diagonal of the matrix to zero, and
calculating the energy levels according to the recursion in
\eqref{free-energy}. The calculations proceed successively through
the upper diagonals: entries at positions
$(1,2),(2,3),...,(n-1,n)$ are calculated first, followed by
entries at positions $(1,3),(2,4),...,(n-2,n)$, and so on. Note
that the entry at $(i,j), j>i$, depends on $\alpha(i,j)$
and the entries at $(i,l)$, $l=i,\ldots,j-1$,
$(l,j)$, $l=i+1,\ldots,n-1$, and $(i+1,j-1)$.
%For Table~\ref{table1}, the entry at position
%$(2,8)$, equal to $-2$ is based on the fact that the base
%at position two of the sequence is $C$, the base at position $8$
%is $G$, and the entry in the table with coordinates $(3,7)$ is
%$-1$. The same value can also be obtained from the observation
%that the table contains $-2$ at position $(3,8)$.

\begin{figure}[tb]
\begin{center}
\epsfig{file=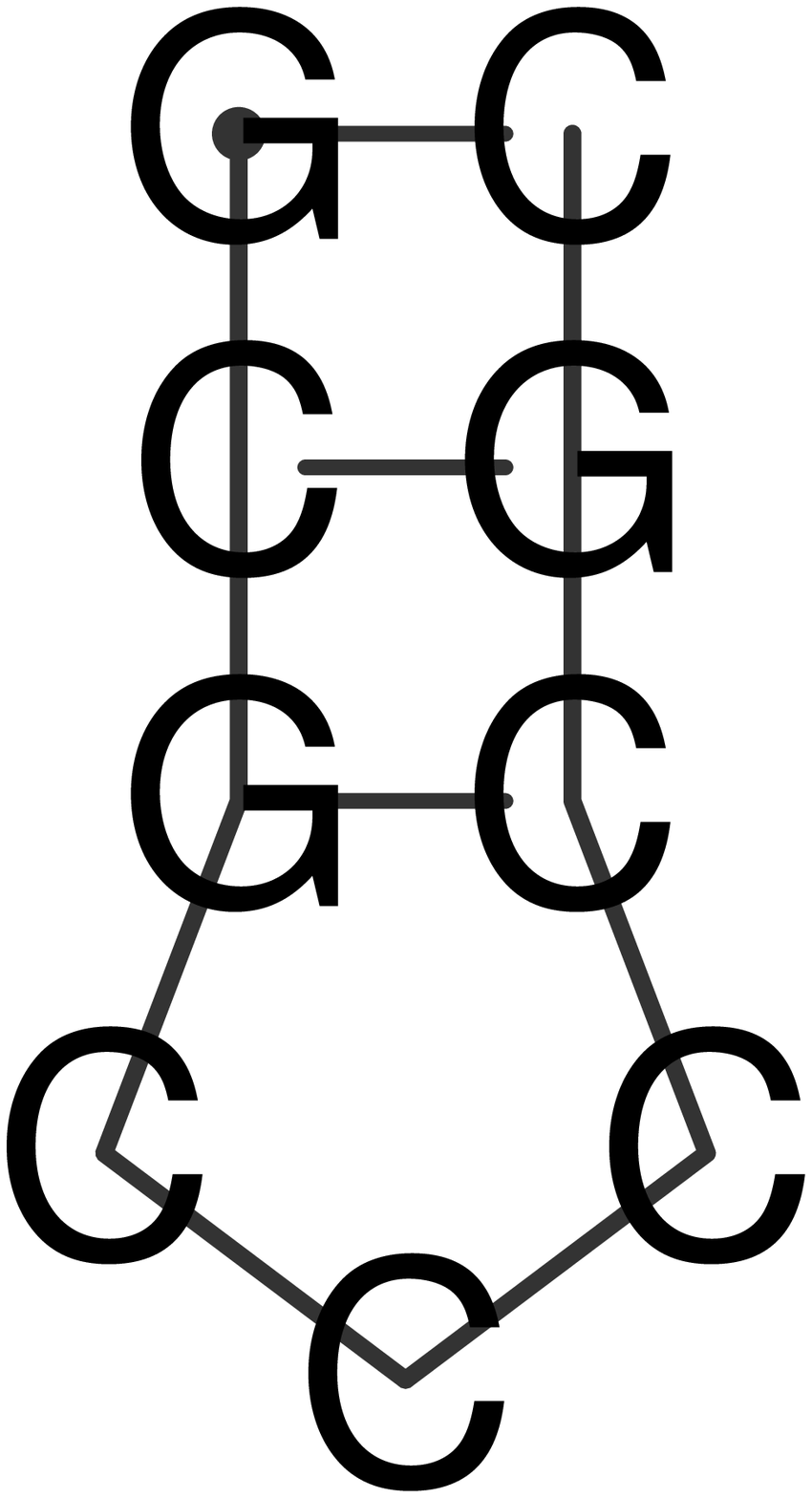, width=1.8cm}
\caption{\label{sec-str1} Secondary structure of the sequence
\bG\bC\bG\bC\bC\bC\bC\bG\bC.}
\end{center}
\end{figure}

\begin{figure}[!t]
\begin{center}
\epsfig{file=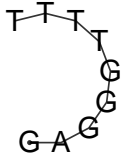, width=1.8cm} \caption{\label{sec-str2}
Absence of secondary structure in the sequence
\bG\bA\bG\bG\bG\bT\bT\bT\bT.}
\end{center}
\end{figure}

The minimum-energy secondary structure itself can be found by the
\emph{backtracking algorithm} \cite{nussinov} which retraces the
steps of Nussinov's algorithm. The secondary structures for the
sequences in Tables~\ref{table1} and \ref{table2},
shown in Figures~\ref{sec-str1} and \ref{sec-str2}, have been found
using the Vienna RNA/DNA secondary structure package
\cite{vienna}, which is based on the Nussinov algorithm,
but which uses more accurate values for the parameters $\alpha(c_i,c_j)$,
as well as more sophisticated prediction methods for base pairing
probabilities.

Tables~\ref{table1} and \ref{table2} show that the minimum free
energy for the sequence {\bG\bC\bG\bC\bC\bC\bC\bG\bC} is $-6$,
while that for the sequence {\bG\bA\bG\bG\bG\bT\bT\bT\bT} is
$-1$.\footnote{Observe that although the free energy in the second
case is $-1$, the sequence is deemed to have no secondary
structure; this is due to the fact that the one possible
complementary base pairing, namely that of {\bA} and {\bT}, forms
too weak a bond for the resultant structure to be stable.} This
fact alone indicates that the number of paired bases in the first
sequence ought to be larger than in the second one, and hence the
former is more likely to have a secondary structure than the
latter.

More generally, one can observe the following characteristics of
free-energy tables: if the first upper diagonal contains a large
number of $-1$ or $-2$ entries, then some of these entries
``percolate'' through to the second upper diagonal, where they get
possibly increased by $-1$ or $-2$ if complementary base pairs are
present at positions $i$ and $i+2$, $1 \leq i \leq n-2$ in the DNA
sequence. The values on the second diagonal, in turn, percolate
through to the third diagonal, and so on. Hence, the free energy
of the DNA sequence depends strongly on the number of non-zero
values present on the first diagonal. This phenomenon was also
observed experimentally in \cite{Breslauer}, where 
%it was shown that transition enthalpies and the free energy can be well
%approximated by the \emph{nearest neighbor} phenomena. There, 
the free energy was modelled by a function of the form
\begin{equation}
E_{1,n}=\kappa+\sum_{i=1}^{n-1} \, \alpha(c_i,c_{i+1}),
 \label{app-energy}
\end{equation}
with $\kappa$ denoting a correction factor which depends on the
number of $G$ and $C$ bases in the sequence \textbf{c}. The
stability of a secondary structure, as well as its melting
properties can be directly inferred from~(\ref{app-energy}). Note
that under the assumption that $\alpha \equiv
-1$ for all pairings, the absolute value of the sum 
in~(\ref{app-energy}) is simply
the total number of pairings of complementary bases between the
DNA codeword \textbf{c} and the codeword shifted one position to
the right or equivalently, the sum of the entries in the first
upper diagonal of the free-energy table. 
These observations imply that a more
accurate model for the free energy should be of the form
\begin{eqnarray*}
\lefteqn{E_{1,n}=\kappa + \gamma_1 \sum_{i=1}^{n-1} \, \alpha(c_i,c_{i+1})} \\
& & \ \ \ \ \  \mbox{} + \gamma_2 \sum_{i=1}^{n-2} \,
\alpha(c_i,c_{i+2})+ \cdots +\gamma_l \sum_{i=1}^{n-l} \,
\alpha(c_i,c_{i+l}),
\end{eqnarray*}
for a correction factor $\kappa$ and some non-zero weighting
factors $\gamma_1 \geq \gamma_2 \geq \ldots \geq \gamma_l$.
Furthermore, the same observation implies that from the
stand-point of designing DNA codewords without secondary
structure, it is desirable to have codewords for which the
respective sums of the elements on the first several diagonals are
either all zero or of some very small absolute value. This
requirement can be rephrased in terms of requiring a DNA sequence
to satisfy a \emph{shift property}, in which a sequence and its
first few shifts have few or no complementary base pairs at the
same positions.
%Another property that can be observed from folding Table I is the
%existence of a path of length three on the ``reverse main
%diagonal'' with entries of the form $-3,-2,-1$ (boldfaced). This
%pattern indicates that the last three bases are paired (in this
%case, to the first three bases). In other words, that path
%indicates that the sequence contains a subsequence and its reverse
%complement (in this case, \bG\bC\bG$\;$ and \bC\bG\bC$\;$). Again,
%the free energy can be seen as an indicator of the presence of
%such sequences. Hence, another constraint governing the possible
%presence of a secondary structure is the \emph{subsequence
%constraint}, implying that a DNA codeword should not
%simultaneously contain a subsequence of length $\geq n$ and its
%reverse complement. As it turns out, both the shift and
%subsequence constraint can be tackled by similar DNA sequence
%design approaches.
%The two coding constraint will be addressed in more detail, and in
%a rigorous mathematical setting, in the next sections of the
%paper.
\section{DNA codewords satisfying a shift property}

In this section, we consider the enumeration and construction of
DNA sequences satisfying certain shift properties, which we shall define
rigorously.

\subsection{Sequence Enumeration}

\begin{definition} The \emph{Watson-Crick (WC) distance} between
two DNA sequences $\mathbf{p} = q_1q_2 \ldots q_n$ and
$\mathbf{q} = p_1p_2 \ldots p_n$ is defined as
\begin{equation}
d_{WC}(\mathbf{p},\mathbf{q})=|\{i: p_i \neq \bar{q_i}\}|,
\end{equation}
\emph{i.e.}, $d_{WC}(\textbf{p},\textbf{q})=
d_{H}(\bp,\bar{\bq})$,
where $d_H$ stands for the standard Hamming distance.

Given a DNA codeword \bq, we shall denote by $\bq_{[i,j]}$,
$1 \leq i<j \leq n$,
the subsequence $q_{i} q_{i+1} \ldots q_j$.
For $0 \leq i \leq n-1$, we define
\begin{equation}
\mu_i(\bq) = n-i-d_{WC}(\bq_{[i+1,n]},\bq_{[1,n-i]}).
%|\{{j:\; 1 \leq j \leq n-i,\;q_j \neq \bar{q}_{i+j}\}}|.
\end{equation}
\end{definition}
\vspace{0.1in} In other words, $\mu_i(\bq)$ is the number of
indices $\ell \in \{1,2,\ldots,n-i\}$ such that $q_{\ell} =
\bar{q_{i+\ell}}$. A shift property of $\bq$ is now simply any
sort of restriction imposed on $\mu_i(\bq)$.

Given $s \geq 1$, let $g_s(n)$ denote the number of sequences, \textbf{q},
of length $n$ for which $\mu_i(\textbf{q})=0$, $i=1,...,s.$
For $n \leq s$, we take $g_s(n)$ to be $g_{n-1}(n)$.

\begin{lemma}
For all $n > 1$, $g_{n-1}(n) = 4 (2^n-1)$.
\label{g_lemma1}
\end{lemma}
\begin{proof}
It is clear that a DNA sequence is counted by $g_{n-1}(n)$ iff
it contains no pair of complementary bases. Such a sequence must
be over one of the alphabets $\{\bA,\bG\}$, $\{\bA,\bC\}$,
$\{\bT,\bG\}$ and $\{\bT,\bC\}$. There are $4 (2^n-1)$ such sequences,
since there are $2^n$ sequences over each of these alphabets, of which
$\bA^n$, $\bT^n$, $\bG^n$ and $\bC^n$ are each counted twice.
\end{proof}

\begin{lemma}
For all $n > s$,
$$
g_s(n)=2g_s(n-1)+g_s(n-s).
$$
\label{g_lemma2}
\end{lemma}
\begin{proof} Let ${\cal G}_s(n)$ denote the set of all sequences
$\bq$ of length $n$ for which $\mu_i(\textbf{q})=0$, $i=1,...,s$.
Thus, $|{\cal G}_s(n)| = g_s(n)$. Note that for any $\bq \in {\cal G}_s(n)$,
$\bq_{[n-s,n]}$ cannot contain a complementary pair of bases, and
hence cannot contain three distinct bases.
Let ${\cal E}(n)$ denote the set of
sequences $q_1...q_n \in {\cal G}_s(n)$ such that
$q_{n-s+1} = q_{n-s+2} = \cdots = q_{n}$, and let ${\cal U}(n) =
{\cal G}_s(n) \setminus {\cal E}(n)$. We thus have
$|{\cal E}(n)| + |{\cal U}(n)|=g_s(n)$.
Each sequence in ${\cal E}(n)$ is obtained from some sequence
$q_1...q_{n-s+1} \in {\cal G}_s(n-s+1)$ by appending $s-1$ bases,
$q_{n-s+2},\ldots,q_n$, all equal to $q_{n-s+1}$.
Hence, $|{\cal E}(n)|=|{\cal G}_s(n-s+1)| = g_s(n-s+1)$,
and therefore, $|{\cal U}(n)|=g_s(n)-g_s(n-s+1)$.

Now, observe that each sequence $q_1 q_2 \ldots q_n \in {\cal G}_s(n)$
is obtained by appending a single base, $q_n$, to some sequence
$q_1 q_2 \ldots q_{n-1} \in {\cal G}_s(n-1)$.
If $q_1 q_2 \ldots q_{n-1}$ is in fact in ${\cal E}(n-1)$,
then there are three choices for
$q_{n}$. Otherwise, if $q_1 q_2 \ldots q_{n-1} \in {\cal U}(n-1)$,
there are only two possible choices for $q_{n}$. Hence,
\begin{eqnarray*}
g_s(n) &=& 3 \, |{\cal E}(n-1)| + 2 \, |{\cal U}(n-1)| \\
&=& 3 \, g_s(n-s) + 2\left(g_s(n-1)-g_s(n-s)\right)
\end{eqnarray*}
This proves the claimed result.
\end{proof}

From Lemmas~\ref{g_lemma1} and \ref{g_lemma2}, we obtain the following
result.

\begin{theorem}
The generating function $G_s(z) = \sum_{z=1}^{\infty} g_s(n) z^{-n}$
is given by
$$
G_s(z) = 4 \cdot \frac{z^{s-1} + z^{z-2} + \cdots + z + 1}{z^s - 2z^{s-1} - 1}.
$$
\label{G_theorem}
\end{theorem}

It can be shown that for $s > 1$, the polynomial $\psi_s(z) = z^s-2z^{s-1}-1$
in the denominator of $G_s(z)$ has a real root, $\rho_s$, in the interval (2,3),
and $s-1$ other roots within the unit circle. It follows
that $g_s(n) \sim \beta_s (\rho_s)^n$ for some constant $\beta_s > 0$. It
is easily seen that $\rho_s$ decreases as $s$ increases, and that
$\lim_{s\rightarrow\infty} \rho_s  = 2$.

\begin{theorem}
The number of length-$n$ DNA sequences $\bq$
such that $\mu_1(\bq) = m$, is $4 \binom{n-1}{m} 3^{n-m-1}$.
\end{theorem}
\begin{proof}
Let $B(n,m)$ be the set of length-$n$ DNA sequences $\bq$
such that $\mu_1(\bq) = m$.
A sequence $\bq = q_1 q_2 \ldots q_n$ is in $B(n,m)$ iff
the set $I = \{i: q_i = \bar{q_{i-1}}\}$ has cardinality $m$.
So, to construct such a sequence, we first arbitrarily pick a
$q_1$ and an $I \subset \{2,3,\ldots,n\}$, $|I| = m$, which can be done
in $4 \binom{n-1}{m}$ ways. The rest of $\bq$ is constructed
recursively: for $i \geq 2$, set $q_i = \bar{q_{i-1}}$ if $i \in I$,
and pick a $q_i \neq \bar{q_{i-1}}$ if $i \notin I$. Thus,
there are 3 choices for each $i \geq 2$, $i \notin I$,
and hence a total of $4 \binom{n-1}{m} 3^{n-m-1}$ sequences $\bq$
in $B(n,m)$.
\end{proof}

The enumeration of DNA sequences satisfying any sort of shift property
becomes considerably more difficult if we bring in the additional
requirement of constant GC-content.

\begin{definition}
The \emph{GC-content}, $w_{GC}(\bq)$, of a DNA sequence
$\bq = q_1 q_2 \ldots q_n$ is defined to be the number of indices
$i$ such that $q_i \in \{\bG,\bC\}$.
\end{definition}

A DNA code in which all codewords have the same GC-content, $w$, is called a
\emph{constant GC-content code}.
The constant GC-content constraint is introduced in order to achieve
parallelized operations on DNA sequences, by assuring similar thermodynamic
characteristics of all codewords. The GC-content usually needs to be in the range of
$30-50\%$ of the length of the code.

The following result can be proved by applying the powerful Goulden-Jackson
method of combinatorial enumeration \cite[Section~2.8]{goulden}.

\begin{theorem}
The number of DNA sequences $\bq$ of length $n$ and GC-content $w$, such
that $\mu_1(\bq) = 0$, is given by the coefficient of $x^ny^w$ in the (formal) power
series expansion of
$$
\Phi(x,y) = {\left(1 - \frac{2x}{1+x} - \frac{2xy}{1+xy}\right)}^{-1}.
$$
\end{theorem}

\subsection{Code Construction}
The problem of constructing DNA codewords obeying some form of a
shift constraint can be reduced to a binary code design problem by
mapping the DNA alphabet onto the set of length-two binary words
as follows:
\begin{equation} \label{map2}
\textbf{A} \to 00, \;\; \textbf{T} \to 01, \;\; \textbf{C} \to 10,
\;\; \textbf{G} \to 11.
\end{equation}

Let \textbf{q} be a DNA sequence. The sequence $b$(\textbf{q})
obtained by applying coordinatewise the mapping given in
\eqref{map2} to \textbf{q} will be called the \emph{binary image}
of \textbf{q}. If $b(\q) = b_0b_1b_2 \ldots b_{2n-1}$, then the
subsequence $e(\q) = b_0b_2 \ldots b_{2n-2}$ will be referred to
as the \emph{even subsequence} of $b(\textbf{q})$, and $o(\q) =
b_1b_3 \ldots b_{2n-1}$ will be called the \emph{odd subsequence}
of $b$(\textbf{q}).
%Thus, for example, for $\q = \mathbf{ACGTCC}$,
%we have $b(\textbf{q})=001011011010$, $e(\q)=011011$ and
%$o(\q)=001100$.

The WC distance, $d_{WC}(\p,\r)$, between two DNA words
$\p,\r$ can be expressed in terms of the even and odd
subsequences, as stated in the lemma below. For notational ease,
given binary words $\x = (x_i)$ and $\y=(y_i)$, we define $\x
\oplus \y = (x_i + y_i)$, the sum being taken modulo-2, and $\x*\y
= (x_iy_i)$.

\begin{lemma} Let \textbf{p} and \textbf{r} be two words of length 
$n$ over the alphabet $Q$, and define 
$\sigma_e = e(\p) \oplus e(\r)$, $\sigma_o =
o(\p) \oplus o(\r)$. Using $\bar{\x}$ to denote the complement of
a binary sequence $\x$, we can express the WC distance between
$\p$ and $\r$ as
$$
d_{WC}(\p,\r)= n - w_H(\bar{\sigma_e}*\sigma_o)
$$
where $w_H(\cdot)$ denotes Hamming weight. Consequently, if
for some length-$n$ DNA sequence $\q$, we take
$\p=\q_{[1,n-i]}$ and $\r=\bq_{[i+1,n]}$, then $\q$
satisfies the $i$-th shift constraint $\mu_i(\q)=0$ iff
\begin{equation}
%\begin{split}
%&
w_H(\bar{\sigma}_{e}*\sigma_{o})=0.
% \;\; i.e. \\
%&b_{2j}+b_{2(j+i)}=b_{n-i}+b_n. \\
%&\text{for}\;\;i=1,...,n-1. \notag
%\end{split}
\end{equation}
\label{wc-binary}
\end{lemma} 

\vspace{0.1in} In a companion paper \cite{milkas}, we described a
sample of construction methods for DNA codes which reduce
undesired hybridization and alow for fully parallel DNA system
operation. Among the constraints identified for this problem are
the base runlength constraint, the constant GC-content constraint,
and the Hamming and reverse-complement Hamming distance
constraint.
%Such constraints can be enforced by
%using well known algebraic codes, involving extended cyclic Goppa
%codes and codes based on generalized Hadamard matrices.
We will show next that it is straightforward to incorporate these
hybridization constraints into a scheme which also allows for
constructing sequences with reduced probability of secondary
structure formation. The idea is based on the use of the non-zero
codewords of cyclic simplex codes \cite[Chapter~8]{hall}
or subsets thereof. Recall that a cyclic simplex code
of dimension $m$ is a simplex code of length $n = 2^m - 1$
composed of the all-zeros codeword and the $n$
distinct cyclic shifts of any non-zero codeword.
%For this purpose, we need the following well known result \cite{hall}.
%\begin{definition}
%A binary sequence of length $2^m - 1$ with the property that each
%nonzero $m$-tuple occurs exactly once in a set of consecutive
%positions (taking cyclic shifts into account as well) is called a
%pseudo-noise sequence of period $2^m-1$ (henceforth, PN-sequence).
%\end{definition}
%PN sequences can be generated by shift registers, and examples of
%such sequences include the codewords of cyclic simplex codes
%$[2^m-1,m,2^{m-1}]$, and their extensions, the first-order
%Reed-Muller codes $[2^m,m+1,2^{m-1}]$.
%Any non-zero
%codeword of such a cyclic simplex code is a PN-sequence of period
%$2^m-1$.
%Consequently, if \textbf{s} is a codeword of the simplex code, and
%\textbf{s}$^i$ is a cyclic shift of \textbf{s} in $0<i<n$
%positions, then \textbf{s} $\oplus$ \textbf{s}$^{i}$=
%\textbf{s}$^{j}$, for some $0<j<n$.

\begin{theorem} Let $\mathcal{C}$ be a DNA code obtained by
choosing the set of non-zero codewords of a cyclic simplex code of length
$n=2^m-1$ for both the even and odd binary code component. Such a
code contains $(2^m-1)^2$ codewords with the property that
for all $i \in \{1,2,\ldots,n-1\}$ and $\bq \in \cC$,
$$\mu_i(\bq) \leq 2^{m-2}.$$
\label{simplex_theorem}
\end{theorem} 
\vspace{0.1in}\begin{proof} It is straightforward to see that for
any $\q \in \cC$, if we let $\p=\q_{[1,n-i]}$, $\r = \q_{[i+1,n]}$, 
then $\sigma_e$ and $\sigma_o$, defined as in Lemma~\ref{wc-binary},
are just truncations of codewords from the simplex
code. Since the simplex code is a constant-weight code, with
minimum distance $2^{m-2}$, each pair
of codewords intersects in exactly $2^{m-2}$ positions. This
implies that there exist exactly $2^{m-2}$ positions for which one
given codeword contains all zeros, and the other codeword contains
all ones. These are the positions that are counted in
$w_H(\bar{\sigma}_e,\sigma_o)$, which proves the claimed
result.
\end{proof}
\begin{example} Consider the previous construction for $m=3$,
and a generating codeword $1110100$. There are $49$ DNA codewords
of length $7$ obtained based on the outlined method. These
codewords have minimum Hamming distance equal to four, and they
also have constant $GC$ content $w=4$. A selected subset of
codewords from this code is listed below.
\begin{equation}
\begin{split}
& \bT\bG\bG\bC\bT\bC\bA,\; \bT\bC\bC\bG\bT\bG\bA,\;
\;\bC\bA\bC\bG\bG\bT\bC,\; \bT\bA\bG\bC\bC\bT\bG,\\
& \bC\bA\bT\bG\bG\bC\bT,\; \bG\bA\bT\bC\bC\bG\bT,\;
\;\bG\bG\bG\bA\bG\bA\bA,\; \bG\bG\bA\bG\bA\bA\bG.\notag
\end{split}
\end{equation}
The last two codewords consist of the bases \bG$\,$ and \bA$\,$
only, and clearly satisfy the shift property with $\mu_i=0$ for
all $i \leq 7$. On the other hand, for the first three codewords
one has $\mu_1=1$, while for the next three codewords it holds
that $\mu_1=2$ (meeting the upper bound in the theorem).
%Furthermore, all these six codewords contain a subsequence of
%length two and its reverse complement.
%An example is given by \bT\bG$\;$ and \bC\bA$\;$ in
%the first codeword.
%But none of the $49$ codewords contains a subsequence of length
%three and its reverse complement. 
Due to the cyclic nature of the
generating code, one can easily generate the Nussinov folding
table for all the codewords \cite{milkas}. Such an evaluation, as
well as the use of the program package Vienna, show that none of
the $49$ codewords exhibits a secondary structure.
%Furthermore,
%the maximum free energy of a codeword is only $-0.06$kcal/mol.
The largest known DNA codes with the parameters specified above
consists of $72$ codewords \cite{gaborit}. This code is generated
by a simulated annealing process which does not allow for simple
secondary structure testing.
\end{example}

%It is
%important to observe that there exist a close connection between
%the codewords of the simplex codes, \emph{pseudorandom} or PN
%sequences, and deBruijn sequences \cite{hall}. Such sequences have
%the property that the minimum Hamming distance of the sequence and
%its cyclic shifts is the largest possible, namely $n+1/2$. This
%property can be used to describe the subsequence properties of the
%codes in Theorem~\ref{simplex}. Furthermore orientable de Bruijn
%sequences \cite{dai} can be used to improve upon the overall
%properties of the construction method.

\end{document}